\documentclass{article}

\usepackage{arxiv}

\usepackage[utf8]{inputenc} 
\usepackage[T1]{fontenc}    
\usepackage{hyperref}       
\usepackage{url}            
\usepackage{booktabs}       
\usepackage{amsfonts}       
\usepackage{nicefrac}       
\usepackage{microtype}      
\usepackage[final]{pdfpages}
\usepackage{graphicx}
\graphicspath{ {./images/} }
\usepackage{times}
\usepackage[round]{natbib}
\usepackage{amsmath}
\newcommand{\mb}{\mathbb}

\title{EB-dynaRE: Real-Time Adjustor for Predicting Brownian Movement Based on a Novel Event-Based Supervised Learning Algorithm}

\author
{Yang Chen,$^{1\ast}$ Emerson Li,$^{1}$\\
\\
\normalsize{$^{1}$Cranbrook Schools, 39221 Woodward Avenue, Bloomfield Hills, MI 48304, USA}\\
\\
\normalsize{$^\ast$To whom correspondence should be addressed; E-mail:  ychen20@cranbrook.edu.}\\
\normalsize{Authors are listed in order of contributions.}
}

\begin{document}
\maketitle
\begin{abstract}
    Stock prices are influenced over time by underlying macroeconomic factors. Jumping out of the box of conventional assumptions about the unpredictability of the market noise, we modeled the changes of stock prices over time through the Markov Decision Process, a discrete stochastic control process that aids decision making in a situation that is partly random. We then did a "Region of Interest" (RoI) Pooling of the stock time-series graphs in order to predict future prices with existing ones. Generative Adversarial Network (GAN) is then used based on a competing pair of supervised learning algorithms, to regenerate future stock price projections on a real-time basis. The supervised learning algorithm used in this research, moreover, is original to this study and will have wider uses. With the ensemble of these algorithms, we are able to identify, to what extent, each specific macroeconomic factor influences the change of the Brownian/random market movement. In addition, our model will have a wider influence on the predictions of other Brownian movements.
\end{abstract}


\section{INTRODUCTION}
Stock prices are influenced over time by a number of underlying macroeconomic influencers. To predict the changes of market prices over time when making investment decisions, financial analysts and major Quant firms came up with their own data analysis algorithms, such as Classification of Market Regimes \citep{levy2003}, to get an edge for their investment.

In fact, what makes predicting stock prices so interesting is exactly its complexities with the turbulent market noises includes the changes of stock prices over time series graphs \citep{chauhan2020} are usually described by Monte Carlo simulation \citep{hastings1970monte}, a method with no closed-form solution (see figure 1). Treated as a Brownian/random movement \citep{roumen2013brownian}, changes of stock prices are predicted by existing algorithms, supervised or unsupervised alike, that work more or less like gambling-achieving a profiting percentage more than the break-even point, 50$\%$, is considered outstanding \citep{taleb2007black}.
\begin{center}
    \includegraphics[width=0.55
    \textwidth]{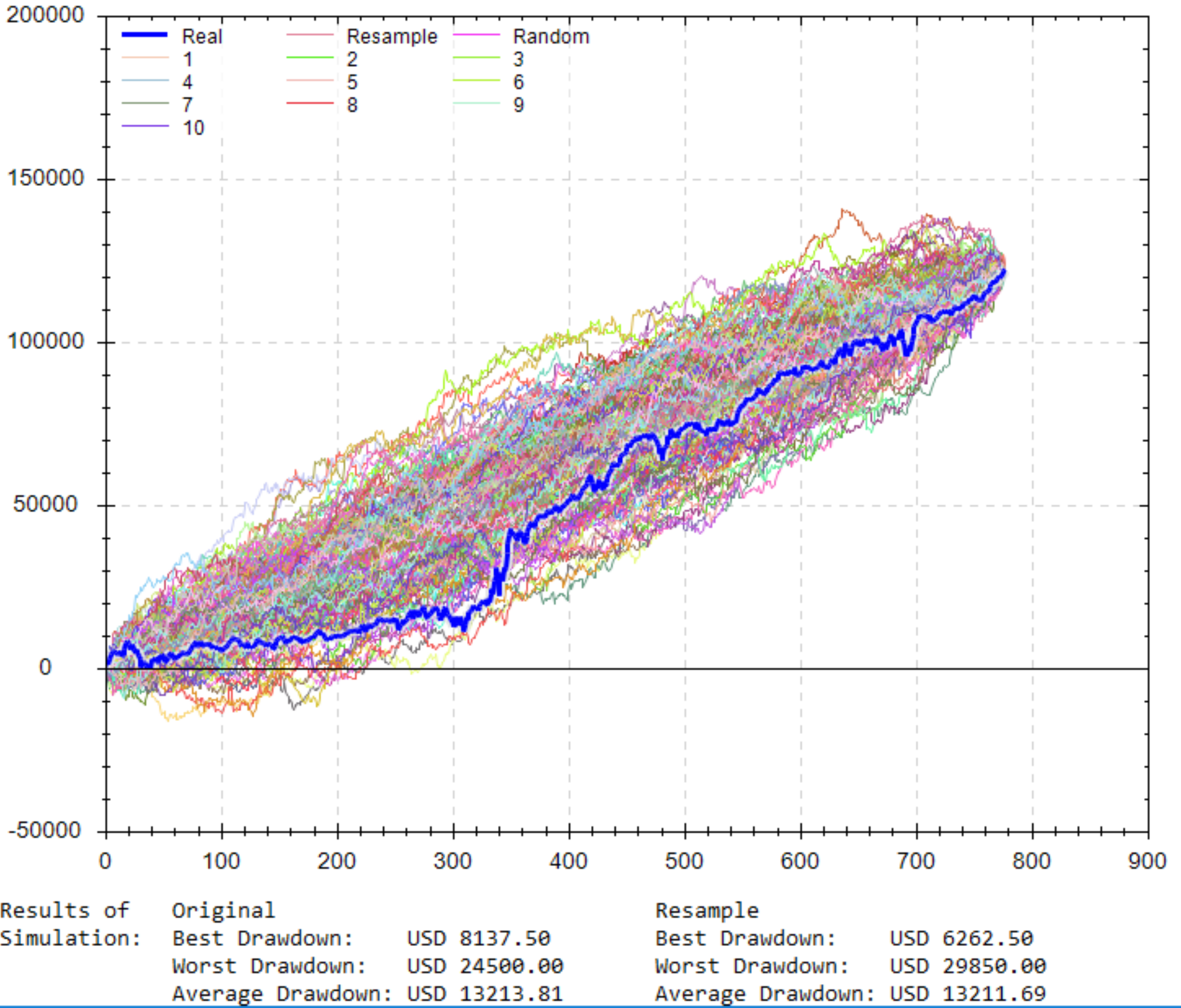}\\
    Figure 1: Monte Carlo Simulation for Trading 
\end{center}
Jumping out of the box of the conventional assumption about the unpredictability of the market noise, the proposed model in this research Event-Based Real-Time Dynamic Regenerative Adjuster (EB-dynaRA) for stock market changes applies emerging generative models to assemble real-time adjustors on top of the existing algorithms based on real macroeconomic events (see a list of them in table 1, measurement for business cycle is shown in Harding et al.'s project \citep{harding2006measurement}). We envision that, because each timely stock price is a result of the Normal Probability Distribution of the previous stock price distribution and Netting Present Value (NPV) \citep{nagalingam1999cim}, we could \textbf{1. }model the changes of stock prices over time through the Markov Decision Process, a discrete stochastic control process that aids decision making in a situation that is partly random; \textbf{2. }then, do a "Region of Interest" (RoI) Pooling of the stock time series graphs in order to predict future prices with existing ones \citep{brinkmann1999art}. This is similar to the idea of active learning proposed in Chen's research in 2019 \citep{chen2020active}, in which the training is gradually specified on mapping regions to reduce computation burden; \textbf{3. }then use Generative Adversarial Network (GAN) \citep{goodfellow2014generative}, based on a competing pair of supervised learning algorithms, to generate future stock price projections on a real-time basis. These supervised learning algorithms are based on a new model novel in this research.
\begin{center}
    \includegraphics[width=0.5
    \textwidth]{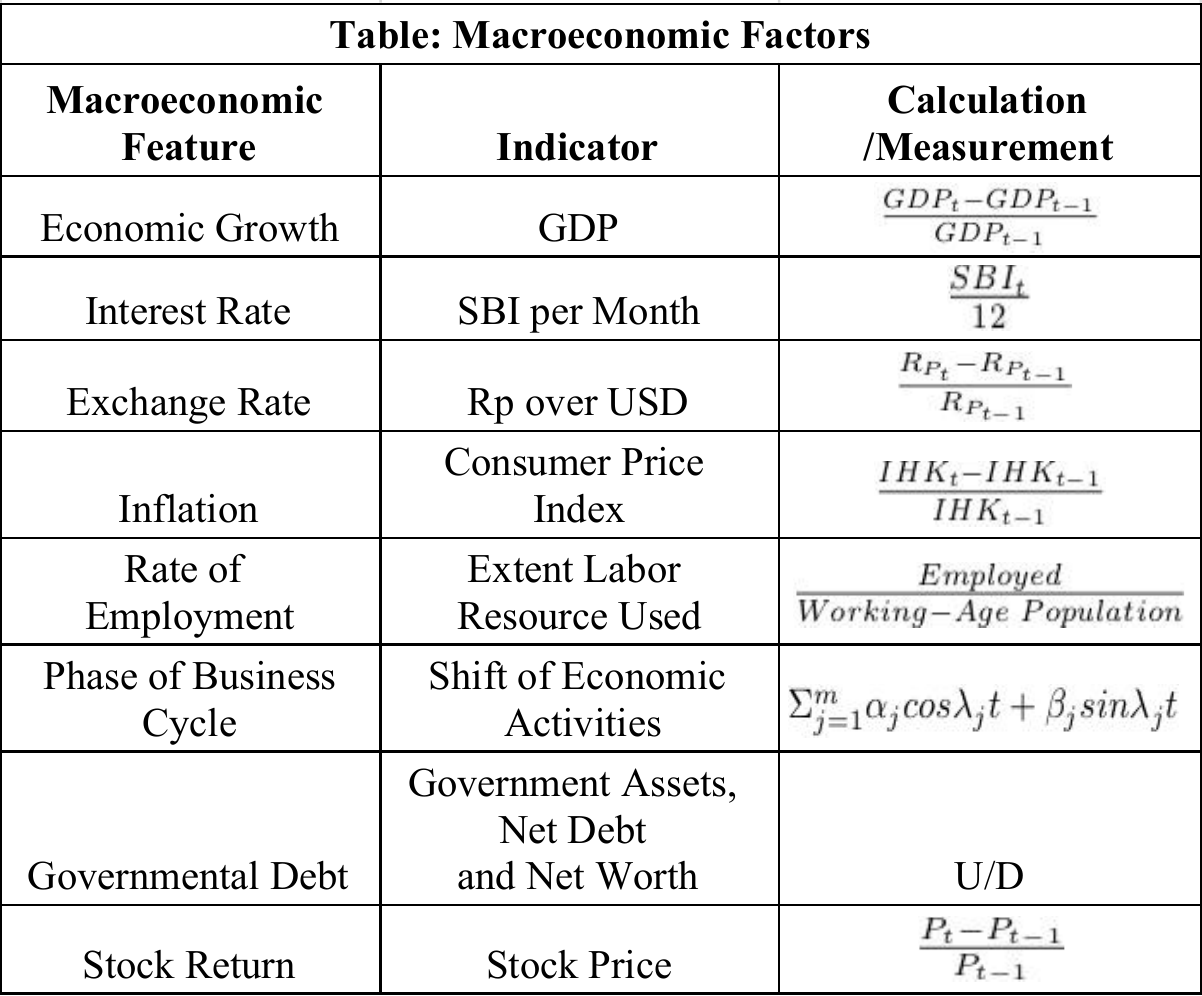}\\
    Table 1: A List of Macroeconomic Factors
\end{center}

\textbf{This research innovates beyond the previous studies in the following ways: }

\textbf{1. }A new mindset to use specific macroeconomic events as parameters to train the supervised learning. This research revolutionized in that it no longer targets the generalized trend of the stock market. Instead, the research breaks down the changes in a stock market into more closely connected industry-based divisions. And then, in each specific division, this project uses quantified event-based parameters to make real-time predictions of the changes of specific stock.

\textbf{2. }A event-based supervised learning algorithm novel to this research. In close collaboration with Dr. Feng Lin, an IEEE Fellow and expert on fuzzy discrete events \citep{lin2002modeling} who is located in Wayne State University, we helped develop and, now, applying the new Fuzzy Discrete Event-Based supervised learning. This learning algorithm is capable of training the supervised model based on events directly -- this is to say, instead of the black box process where all the parameters do not possess intrinsic meanings when being trained (see table 1 below as a summary, updated based on \citep{dong2019}), the algorithm in this research explicitly demonstrates the learning process and the parameters being learned.

\textbf{3. }A cutting edge result. Not only are all the learning parameters explicitly displayed as a part of the training, the model in this research is able to achieve cutting-edge result when predicting stock market changes because of this newly quantified direct correlation between events and their impacts. Once a larger database is able to be built with the trained information of how a specific event would contribute to the stock market, the model in this research will be able to predict changes of any industry-specific stock accurately. Note that this presented model works better when the applications are industry-specific because it has been researched and confirmed of such in Shynkevich's work in 2015 \citep{shynkevich2015stock}.

\begin{center}
    \includegraphics[width=1 \textwidth]{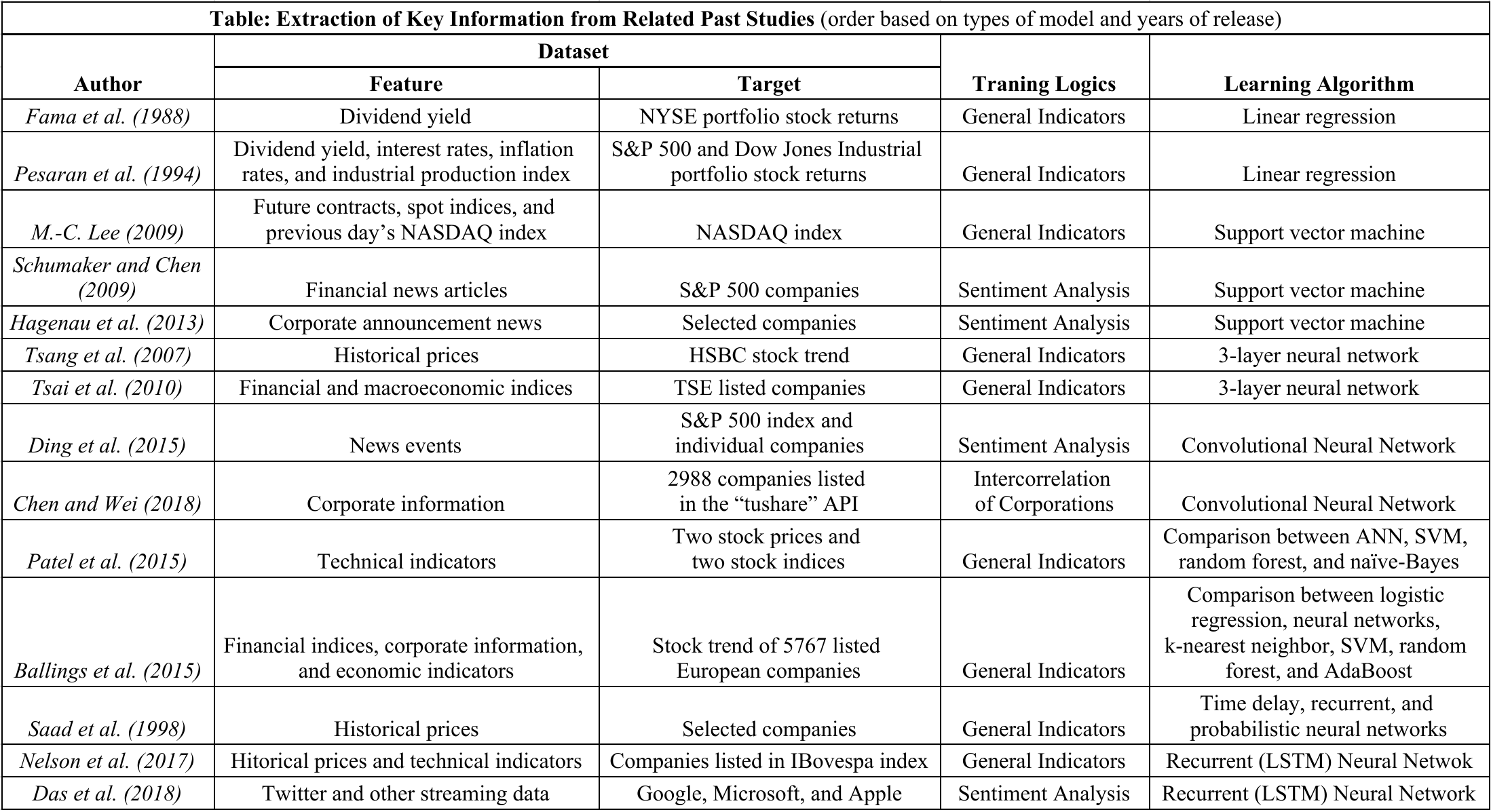}
    Table 2: Extraction of Key Information from Related Past Studies
\end{center}
\section{RELATED WORKS}

\subsection{Traditional Non-Computational Stock Market Prediction Methods:}
Stock exchanges have long been a way for people to invest and profit since as early as Dutch began practicing joint-stock as a business model since the year 1602. In order to make a profit, the simple rule is to approximate the return of the investment with $I = \Delta P = P’ - P_0$ ($I$ being interest, $P$ being the price, $P'$ being the future sell price, and $P_0$ being past buy-in price), adjusted with other economic indicators such as inflation within the time span. Stock strategies, more or less, are to maximize interests through predicting trends of invested stocks and apply appropriate investment strategies. 

The history of stock investments extends way beyond when there is the computation available for making quantitative investment strategies. As a result, many human-intuition-based methods are used throughout the stock history with a number of technical metrics as references for strategy making.

\subsubsection{Profit:}
The traditional definition for success in stock investment (or any kind of investment in general) Price-to-Earings Ratio (P/E) is an important measure. Its value is expressed using the ratio of what investors pay to their prospective earnings. The P/E is calculated by taking the current stock price and dividing it by the earnings per share. If this value is low, it more often than not means a bargain price (or simply indicates that a specific investment lacks intrinsic return prospect). This value, however, similar to many other technical indicators, is used more fairly when is only applied to compare between companies that are in the same/similar industry(ies).

\subsubsection{Intuitive Investment:}
Generally speaking, investors in history would believe in an overall pattern of the changes in the stock market. This is to say that either the price of stock astronomically drops and rises in a short amount of time, then investors would think that the market would make its own “adjustments” and these “adjustments” are predictable and, thus, potential for profiting. If a stock rises rapidly, then investors would more likely than not decide not to invest in the stock or would sell off what they have. The reason is investors believe in extreme cases where a stock rises rapidly, then there is the chance that the market will make its adjustment so that the price will drop as rapidly in a short time. On the other hand, the investors would also prefer not to buy in or sell out whatever stock they have that is rapidly falling, because of their fears of continued deteriorations.

There are a couple of technical metrics people may pertain to to adjust their investment decisions:

\textbf{\textit{Momentum}} According to a work published in 1993 \citep{titman1993}, stocks' performances have inertia, for better or worse. However, this influence is only demonstrated in a short run. Jegadeesh et al. discusses that the general strategies to buy stocks that exhibit good performances (winners) and sell stocks that exhibit bad performances (losers) express positive outcomes so long within a holding period of 3-12 months. Within this short-term period, stocks generally delay its reactions to macroeconomic factors. 

Because of the different legislature and policies across different time and countries, there are variations of the short-term momentum that could potentially exploited. For example, Lehmann states in his 1990 research that, despite the intensive transaction for shot-term buy and sell, frequent trading based on information from the past days and weeks can result in abnormal returns \citep{lehmann1990}. Similarly, as an application of Jegadeesh et al.'s study, Lin et al. uses the artificial intelligence technologies that are lately available as a result of the recent boom of computational growth to take advantage of the short-term momentum \citep{Lin2019}. In the Chinese market, specifically, information disparity ends up with huge short-term trading opportunities for either parties with excessive media information or major corporations that are capable of exploiting public information. Lin et al.'s work makes use of these media coverages as a way to predict turning points of a stock from negative to positive return and make corresponding investment decisions, with risk adjusted.

However, in the long-run, momentum has diminishing effects, which leads to a long-term uncertainty, as a conclusion drawn from Jegadeesh's extensive survey and research \citep{titman1993}. For example, many winner stocks that perform well for the first 12 months of holding exhibit about $10\%$ return rates and suffer from continuous losses for the 24 months that follow. This pattern discovered in Jedadeesh et al.'s work leads to debate about predicting long-term investment based on past stock performances.

\textbf{\textit{Mean Reversion}} Balvers et al. discusses in their research an extension to the short-term momentum -- the long-term pattern on top of the short-term momentum \citep{Balvers2000}. Provided that the effects of momentum are limited to a short period of time \citep{titman1993}, the market would make its correction and neutralization in a long run. Surveying through the stock market of 18 countries from 1969 to 1996, Balvers et al. discovered strong evidence of mean reversion by which the prices of individual stocks will eventually return to a mean value over a sufficient time horizon, be it a winner or loser within a short period of time. Their research determines that the half-life of such reversion is three to three and half years based on huge amount of data based on past market index \citep{Balvers2000}. The result of their studies have been widely used for making contrarian investment strategies.

\textbf{\textit{Martigales}} A martingale is a statistical term that describes any sequence of random variables such that the conditional expectation of the next value at a given time in the sequence equates the present value given all previous values \citep{Balsara1992}. Economists who attach to this idea theorize that past price of a given stock should have no effect over the future price in an efficient market (the only reason it does is because none of the markets in the world today is totally efficient). This statistical idea is widely adopted in today's market assumption which states that the accurate estimate of the stock market should follow a Brownian/random movement. As a result, many of the quantitative trading strategies more or less treat the predicting the time series of stock market changes over time as gambling. However, we believe in this research that many macroeconomic factors could influence the changes of the stock market in a short run as Lehmann and Lin et al. show in their research to exploit such short-term effects of macroeconomic effects \citep{lehmann1990} \citep{Lin2019}.

This idea, also, is important for the development of the presented research here, because of the strong evidence that states the best estimate of the near-future changes of the stock market is based on today's price. Together with Jegadeesh et al.'s conclusion that short-term events have diminishing effects over a long period of time, we device a rolling-window model detailed in section 3.3.

\textbf{\textit{Searching for Value}} According to Eugene Fama and Kenneth French's famous and Nobel-Price-winning Fama-French Three-Factor Model, the returns of investments in asset pricing and portfolio management can be described using three factors, namely market risk, the outperformance of $\frac{small}{big companies}$, and the outperformance of $\frac{high\ book}{market}$ versus $\frac{small\ book}{market}$ \citep{fama1996}. Later, Fama and French also added profitability and investment factors in 2005 and came up with a Five-Factor Model to evaluate asset pricing \citep{fama2015}. This model is also particularly useful to consider when predicting to changes of the stock prices, as it, on a certain level, determines the pricing of stocks.

\subsection{Stock Market Prediction Methods Based on Machine Learning:}
Chen et al. from Fundan University School of Data Science implemented a \textbf{\textit{graph convolutional neural network (GCN)}} to incorporate information of related corporations for a target company’s stock price prediction. They first create a graph of related corporations based on the target company’s investment information \citep{chen2018incorporating}. Their two approaches to implementing the information from related corporations were based on a pipeline model and a joint model. The financial market is said to be informationally efficient, therefore based on the Efficient Market Hypothesis (EMH), it would be intuitive to venture that a change in stock price of a target corporation would be affected by related corporations. Based on this work’s experiential results, incorporating related corporations improved the stock prediction accuracy. However, one of the two future directions to explore is to factor in heterogeneous information sources for better prediction, rather than solely numerical information which this paper focussed on.

Taking one stape further from directly finding mapping through neural networks, Dong from Middlesex School created the framework \textbf{\textit{Dynamic Advisor-Based Ensemble (dynABE)}}, which examines a given company’s areas of interest to diversify the features by creating advisors for each of these areas \citep{dong2019}. DynABE’s four aims are to incorporate domain-specific information based on a given company’s areas of interest, to implement a first-level ensemble learning framework (for each advisor), to establish a second-level ensemble learning framework by combining the advisors, and to construct an effective online learning model. Dong chose critical metal companies as the case study for dynABE. For base models to combine using different ensemble methods, Dong used linear regression, logistic regression, SVM, XGBoost, and Rotation Forest. He then put them into an ensemble using stacking. Finally, the method for combining advisors into the second-level ensemble is an online update strategy during the active trading period. Through experimentation, dynABE outperformed baseline models of SVM, neural network, and random forest. However, one possible area of elaboration of dynABE is that this model was only tested for one industry, critical metals, thus allowing Dong to implement principal component analysis (PCA) to establish the advisors. Critical metals is an industry for which the features are highly correlated, allowing PCA to be run. The scalability of this model is problematic in that not every industry would have as highly correlated features as the critical metal industry.

Dr. David Nelson from Minas Gerais Federal University’s department of computer science implemented \textbf{textit{Long Short-Term Memory neural networks}} to predict future trends of stocks as an additional preprosessing of the input information. An LSTM network’s distinction is that LSTM has feedback connections rather than standard feedforward neural network, therefore allowing for processing of sequences of data rather than singular data points. Nelson utilized this functionality to analyze stock price history to predict future trends \citep{nelson2017stock}. LSTM networks have been the most effective algorithms for Natural Language Processing, and have thus been used to analyze news text data to predict stock prices.  In conclusion, Nelson’s LTSM network outperformed other machine learning models despite the input dimension being very large and relatively high variance. This application of LTSM is similar to our proposed model in that both implement work in a rolling window fashion, for which a database, perhaps candlestick, like Nelson’s, can be pre-processed into a time-series graph with each different point in the graph having unique weights.

There are also efforts trying to detect sentiment information and use it as input to predict stock market changes based on direct media/online information. Dr. Sushree Das of Indian National Institute of Technology Department of Computer Science and Engineering analyzed Twitter streaming data, a continuous flow of data collected in real-time \citep{das2018real}. Thus, with every iteration of data influx, a model would become more accurate, reshaping the model prepared from historical data. According to the Martingale principle, recent stock data is significantly more influential than historical stock data, therefore analyzing streaming data should maximize accuracy. Das performed a times series analysis with sentiment analysis (input data is classified as having a positive or negative sentiment) through data collected with the Twitter Streaming API. The sentiment analysis was performed using recurrent neural network (RNN) components. Das’s model of using \textbf{\textit{sentiment analysis}} to predict future stock prices using time-series inputs is similar to our proposal, but with sentiment analysis rather than an event-based dynamic regenerative adjustor. Through experimentation, we would like to compare the effectiveness of these two methods.

Last but not the least, \textbf{\textit{Monte Carlo simulations}} are the most widely used non-machine-learning computational method to make predictions for stock market changes, assuming it more or less as a random movement. Monte Carlo simulations are used to predict the uncertainty and risk associated with prediction models \citep{kroese2014monte}. Although the Monte Carlo method can be applied to models in a variety of fields, we will discuss its applications to finance. The primary issue with the Monte Carlo method in finance is that it assumes a perfectly efficient market while ignoring macroeconomics events and cyclical factors. Our proposed model aims to address this issue through setting up an ensemble model constructed in a way such that the first-level ensemble is an LSTM neural network which receives a processed time-series graph input comprising the last ten days of stock data, with each hidden layer outputting an event matrix rather than the weights in a typical black box network to be saved for use in the second-level ensemble which is a generative adversarial network (GAN) incorporating the event matrices from the first level.
\section{Methodology:}
In this section, the goal is to give a detailed description of EB-dynaRA after the event-specific matrices have been formulated. To pre-process the time-series data to begin with so it is ready to be trained, we first apply Region of Interest to pool the information and extract key features from time-series graphs. Incorporating this process to preprocess image data like the stock graphs is similar to the ideology of active learning mentioned in Chen's paper \citep{chen2020active}, which would, in effect, reduce computation burden by focusing on only the regions providing the most information. Then, we use Generative Adversarial Networks (GAN) to produce real-time adjusters for traditional models of predicting stock market changes based on macroeconomic events. To make these macroeconomic events directly trainable in the forms of parameters, we introduce a novel model in collaboration with Dr. Feng Lin -- a supervised learning algorithm based on Fuzzy Discrete Event System.

To begin with, we need to find stocks that are suitable for our training purpose. To do this, we select only equities that have pairs whose absolute values of correlations are higher than 0.95. To find this value of correlation, we used permutation test \citep{fisher1936design}, which is a test of correlation based on resampling. Randomizing for a large number of trials (we took $10^4$, which is still relatively fast for computers to do), permutation test looks at the chances of the observed situations to happen. The chances being lower than $5\%$ which would be sufficient to reject the null hypothesis represent the correlation higher than $95\%$, which correspondingly suggests statistical significance for correlations.
\begin{center}
    \includegraphics[width=0.6
    \textwidth]{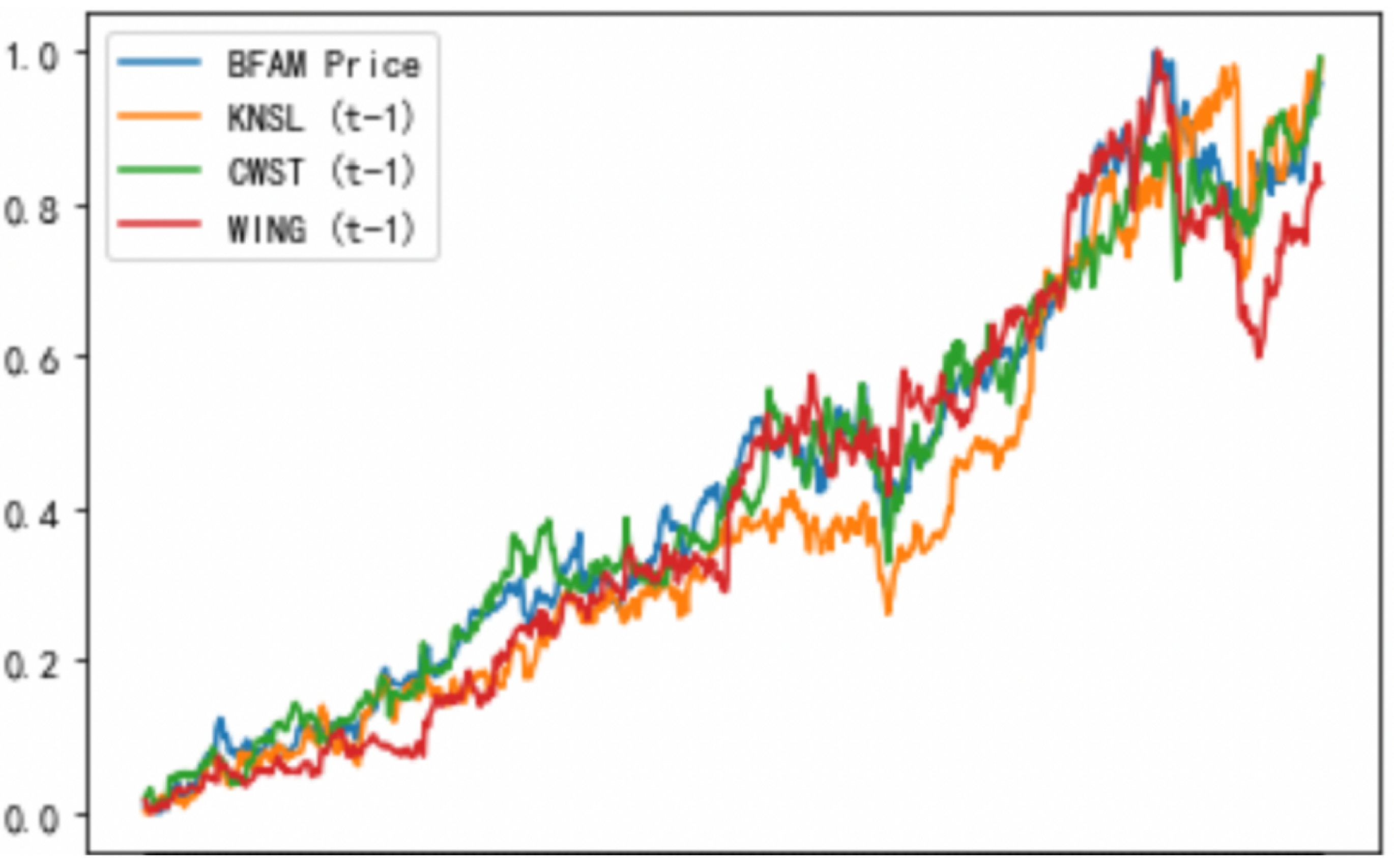}\\
    Figure 2: Pair-Trading Signals
\end{center}
Based on our standard above, we chose the following 88 equities:
'QURE', 'PLNT', 'REXR', 'HAE', 'ASND', 'MR', 'BFAM', 'DNKN', 'KOPN', 'HUBS', 'CWH', 'AVGR', 'MANT', 'RGEN', 'DECK', 'CACC', 'ZEN', 'ATTO', 'PRO', 'ADRO', 'ACRS', 'WDFC', 'JAGX', 'DSKE', 'EVTC', 'AZPN', 'LII', 'CAMT', 'APPF', 'CYBR', 'QTWO', 'MORN', 'STRA', 'ICFI', 'LPTH', 'SFUN', 'GLOB', 'RRTS', 'DLPH', 'HCR', 'TRHC', 'ENTG', 'MATW', 'OBLN', 'BOOT', 'CDXS', 'NVCR', 'RWLK', 'CWST', 'SND', 'EVBG', 'SLCA', 'INPX', 'CHGG', 'AUDC', 'GWRE', 'ROLL', 'FET', 'CLB', 'FIVN', 'WK', 'INXN', 'ENPH', 'PCTY', 'WIX', 'MGEN', 'TTEK', 'TPB', 'CLIR', 'JYNT', 'MGRC', 'EHTH', 'MODN', 'LPSN', 'GNRC', 'GLG', 'BKI', 'POOL', 'CROX', 'KNSL', 'EMKR', 'SEDG', 'INS', 'OGS', 'PULM', 'AAOI', 'WING', 'RPD'

\subsection{Region of Interest (RoI):}
Region of Interest (RoI) has been frequently used as a way to extract key information/key area of training out of images being analyzed \citep{brinkmann1999art}. It is particularly useful in Medical Imaging, the process of which consists scanning an entire area and abstract information about a specific area where the abnormality would potentially be.

Similarly, this model could as well be applied to time-series graph of the stock market to extract significant information directly so that computation resources will not be wasted on distracting information such as the background. The way it is done is similar to what is shown in figure 3 below.

\begin{center}
    \includegraphics[width=0.6
    \textwidth]{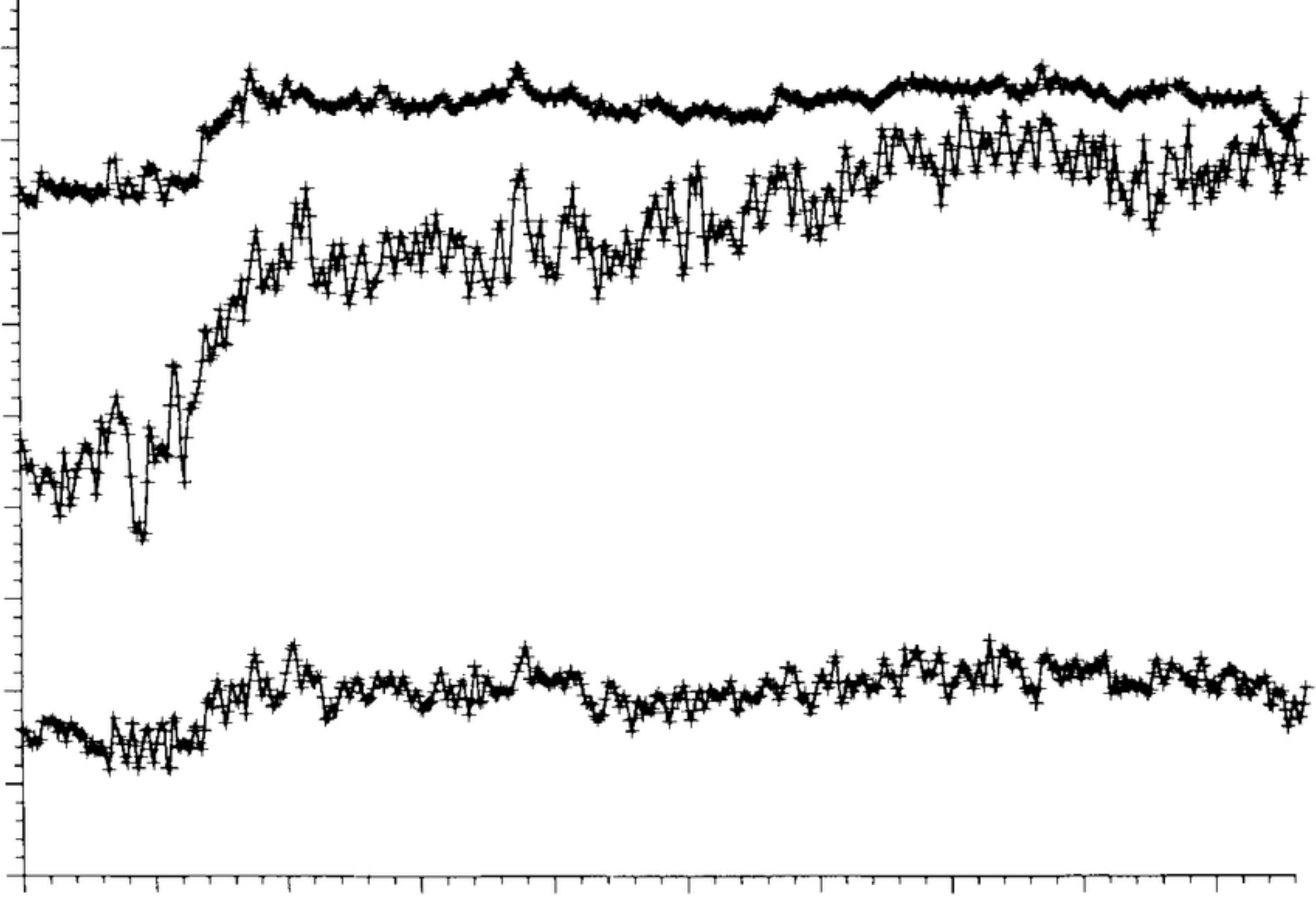}\\
    Figure 3: Region-of-Interest (RoI) Extracts Function Fluctuations in Terms of Time
\end{center}

We use this method to extract key information of all the stocks for the convenience and speed of training in 3.2 and 3.3.

\subsection{Generative Adversarial Networks (GANs):}
\begin{center}
    \includegraphics[width=0.7
    \textwidth]{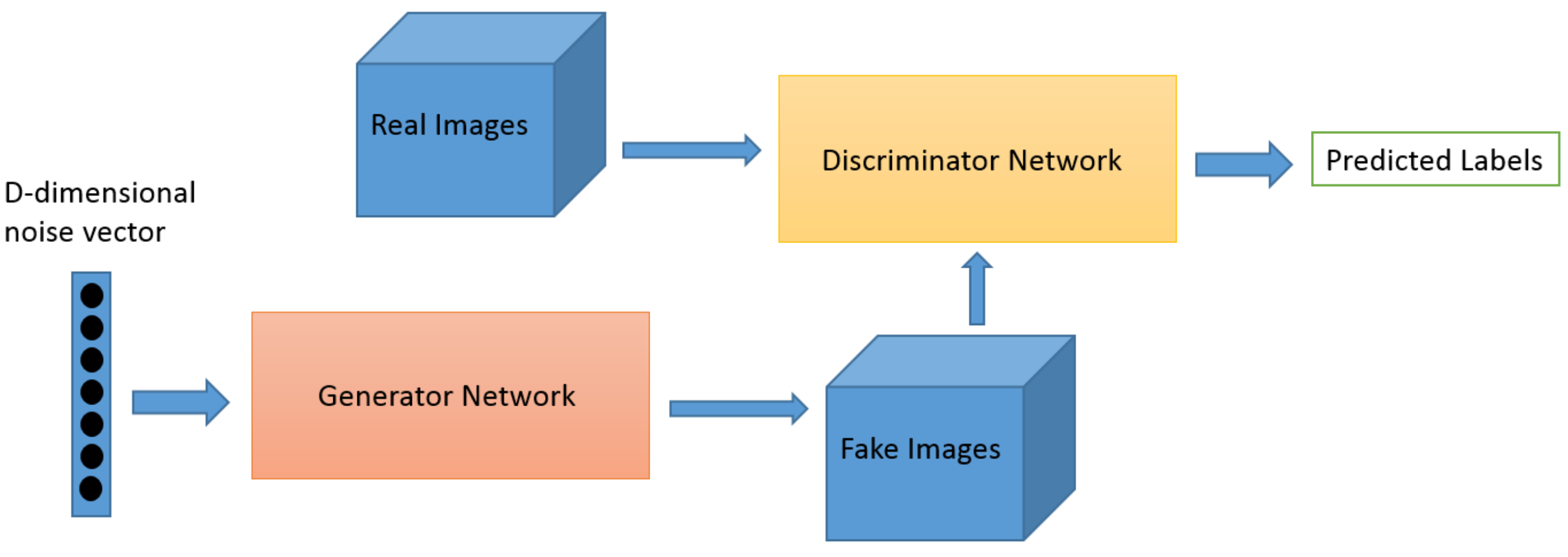}\\
    
    \includegraphics[width=0.7
    \textwidth]{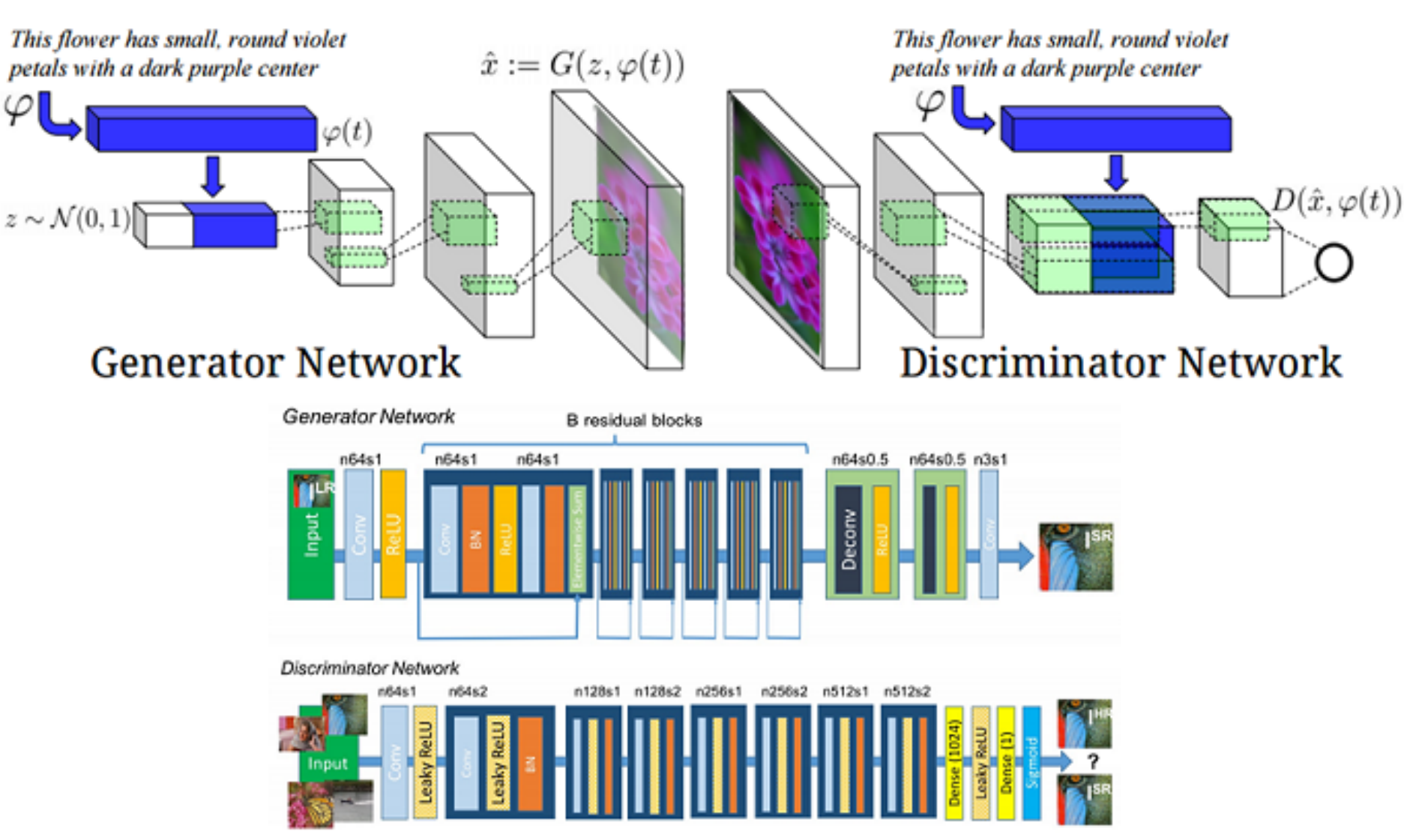}\\
    Figure 4: GANs Mechanism (\textbf{Top:} Summary;\textbf{Bottom:} More Details)
\end{center}
Generative Adversarial Nets (GANs) is a very recent model developed by Goodfellow et al \citep{goodfellow2014generative}. Their works show some of the most cutting-edge results for regenerating things that resemble the training samples to a high quality. For example, with a large number of images of cats, GANs is capable of generating new pictures of cats.

This model becomes particularly useful for our purpose, because our objective is to find the most useful features, called macroeconomic factors, from a large number of stocks and we hope to regenerate the effects of these macroeconomic events based with GANs

Figure 4 above shows the mechanisms how GANs work. A generator network train based on a set of real images and try to produce new images that are similar enough to the real original images so that it can fake the discriminator network, which is also trained based on the set of real images. Maximum likelihood is a way to define a model that provides an estimate of a probability distribution, parameterized by parameters $theta$. Then, the likelihood can be calculated with $\Pi^m_{i=1}P_{model}(X^{(i)};\theta)$, for any data set with m samples of training data $x^{(i)}$. The result of the competing style of the training shown above is expressed below in figure 5.
\begin{center}
    \includegraphics[width=0.7
    \textwidth]{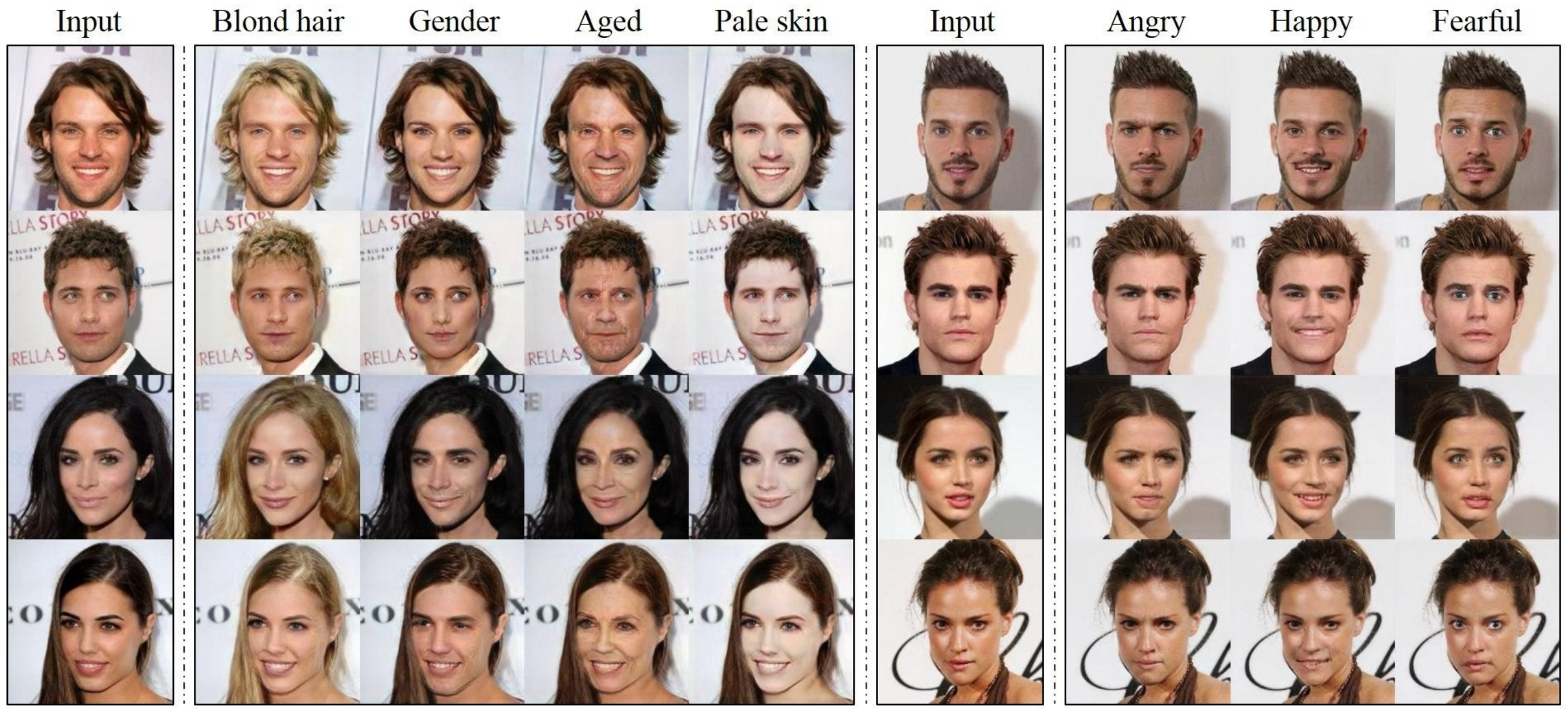}\\
    Figure 5: GANs Results
\end{center}
In this study, the generator network is formed by a new supervised learning algorithm developed with the help of Dr. Feng Lin, the detail of which is elaborated below.

\subsection{Supervised Learning of Fuzzy Discrete Event Systems by Backpropagation:}
The key difference and innovation in this model is the new supervised learning algorithm that is used in this presented research. This model is different from traditional supervised learning that our parameters that we are training with have intrinsic meanings, unlike the black-box process the usual training do (people generally lack the knowledge of what each parameter mean and how the machine gets there).

The way how the novel Event-Based supervised learning works is elaborated in 3.3.1 (since this model is new, this paper will explain about it in details). To make sure our model aligns with the Martingale economists' belief that the present is the most significant basis for the immediate future price and that future is independent from the past given the present.

\subsubsection{Novel Supervised Learning Algorithm:}
\textbf{\textit{Fuzzy Discrete Event Systems}} A fuzzy discrete event system is of the following form: $$\bar{G} = (\bar{Q}, \bar{\Sigma}, \bar{\delta}, \bar{q_0})$$

$\bar{Q}$ is a matrix space of the shape $[0,1]^N$ that represents the state of a certain thing. In the case of this presented research, this state is the value of a given stock price. $N$ is the dimension of the state space (which forms $N$-dimensional row vectors). In the beginning, the system starts at the state $\bar{q_0}$. After a certain series of fuzzy discrete events (in our case, macroeconomic events) happens, let us call these events $\bar{\Sigma}$, the system would change from one state to another. We can specify a single event as $\bar{\delta}: \bar{Q}\times \bar{\Sigma} = \bar{Q}$. We redefine what happens as $\bar{\delta}(\bar{q}, \bar{\sigma})=\bar{q}\circ \bar{\sigma}$. In this case, $\circ$ stands for a fuzzy reasoning operator. For more details about how fuzzy discrete system works, see Lin et al's research \citep{lin2002modeling}.

Now we define two key components, the fuzzy state vector $\bar{q}$ as 
\[
   \bar{q_0}=
  \left[ {\begin{array}{cccc}
   S_1 & S_2 & ... & S_N \\
  \end{array} } \right]
\],
and the elements in the fuzzy event matrix $\bar{\sigma}$ as 
\[
   \bar{\sigma}=
  \left[ {\begin{array}{cccc}
   a_{1,1} & a_{1,2} & ... & a_{1,N} \\
   a_{2,1} & a_{2,2} & ... & a_{2,N} \\
   ...\\
   a_{N,1} & a_{N,2} & ... & a_{N,N} \\
  \end{array} } \right]
\].
With the two components above, we can calculate $\bar{q_1}$ based on $\bar{q_0}$ and some event $\bar{\sigma_1}$ through the following calculation:
$$
\begin{aligned}
\bar{q_1} 
    &= \bar{\delta}(\bar{q_0}, \bar{\sigma_1})\\
    &= \bar{q_0}\circ \bar{\sigma_1}\\
    &= {\left[ {\begin{array}{cccc}
   S_1 & S_2 & ... & S_N \\
  \end{array} } \right] \circ
  \left[ {\begin{array}{cccc}
   a_{1,1} & a_{1,2} & ... & a_{1,N} \\
   a_{2,1} & a_{2,2} & ... & a_{2,N} \\
   ...\\
   a_{N,1} & a_{N,2} & ... & a_{N,N} \\
  \end{array} } \right]}\\
  &= {\left[ {\begin{array}{cccc}
    {max\{S_1\times a_{1,1}, ..., S_N\times a_{N,1}\}} & {...} & {max\{S_1\times a_{1,N}, ..., S_N\times a_{N,N}\}}
  \end{array} } \right]}\\
\end{aligned}
$$
This part above shows how to calculate between states and events when both are known. The following part will discuss how to find the fuzzy discrete event matrices, given a large number of data, without knowing the initial or the final state.

\textbf{\textit{Supervised Learning of FDES}} We assume that a sequence of fuzzy discrete events $\bar{\sigma_1},\bar{\sigma_2},...,\bar{\sigma_N}\in \mb{R},[0,1]$ occur in a fuzzy discrete event system $\bar{G}$ as shown below in figure 4 below:
\begin{center}
    \includegraphics[width=0.8
    \textwidth]{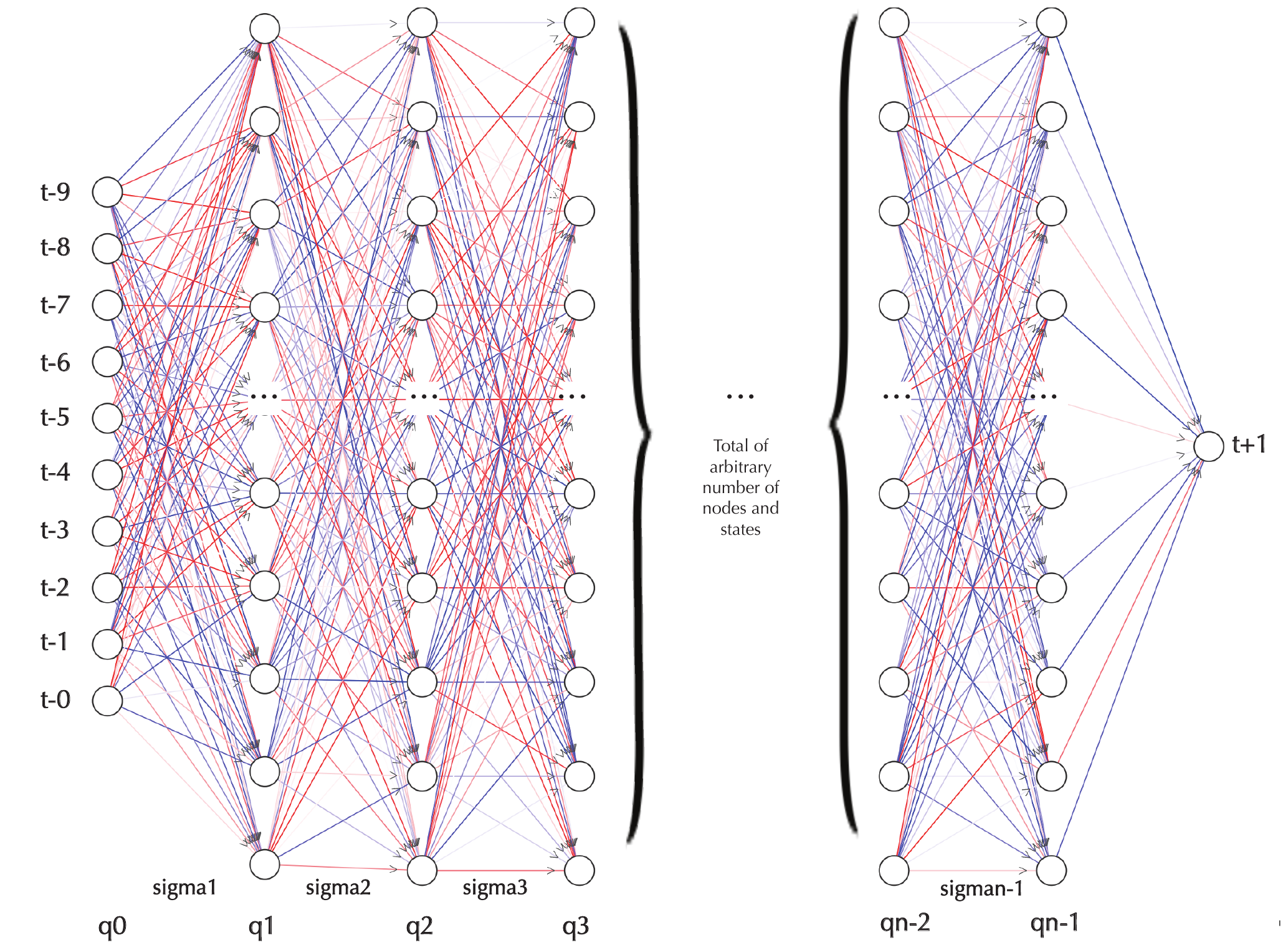}\\
    Figure 6: Training Architecture
\end{center}
As shown above, each layer denotes the fuzzy state $$\bar{q_n}=
\left[ {\begin{array}{cccc}
   S^n_1 & S^n_2 & ... & S^n_N
  \end{array} } \right]
$$ 
and so on. Similarly, each connection between two adjacent layers (a.k.a. the activation function) are defined as the fuzzy event matrices:
$$\bar{\sigma_k}=
\left[ {\begin{array}{cccc}
   a^k_{1,1} & a^k_{1,2} & ... & a^k_{1,N}\\
   a^k_{2,1} & a^k_{2,2} & ... & a^k_{2,N}\\
   ...\\
   a^k_{N,1} & a^k_{N,2} & ... & a^k_{N,N}\\
  \end{array} } \right]
$$
The calculation to get $\bar{q_k}$ is similar to the general formula to calculate the next event and is expressed as:
$$
\begin{aligned}
\bar{q_{n+1}} 
    &= \bar{\delta}(\bar{q_n}, \bar{\sigma_{n+1}})\\
    &= {\left[ {\begin{array}{cccc}
   S^n_1 & S^n_2 & ... & S^n_N \\
  \end{array} } \right] \circ
  \left[ {\begin{array}{cccc}
   a^{n+1}_{1,1} & a^{n+1}_{1,2} & ... & a^{n+1}_{1,N} \\
   a^{n+1}_{2,1} & a^{n+1}_{2,2} & ... & a^{n+1}_{2,N} \\
   ...\\
   a^{n+1}_{N,1} & a^{n+1}_{N,2} & ... & a^{n+1}_{N,N} \\
  \end{array} } \right]}\\
  &=\left[ {\begin{array}{cccc}
   S^{n+1}_1 & S^{n+1}_2 & ... & S^{n+1}_N \\
  \end{array} } \right]
\end{aligned}
$$
Similar to the logic of maxpooling, we define $\circ$ as the function for max-product, since it is test proven to be a sufficient estimate as long as kept consistent. Therefore,
$$
S^{n+1}_j = max\{S^{k-1}_1 a^{k}_{1,j}, S^{k-1}_2 a^{k}_{2,j}, ..., S^{k-1}_N a^{k}_{N,j}\}
$$
This function is not differentiable, though. So to make it fit for being an activation function, we use another exponential function to approximate the max function:
$$
\begin{aligned}
S^{n+1}_j 
    &= max\{S^{k-1}_1 a^{k}_{1,j}, S^{k-1}_2 a^{k}_{2,j}, ..., S^{k-1}_N a^{k}_{N,j}\}\\
    &\approx \frac{1}{\delta}ln(\sigma^N_{m=1}e^{\delta S_m^n a^{k+1}_{m,j}})
\end{aligned}
$$
, in which the $\delta$ is a parameter that is trainable by the supervised learning process.

Now, say we have a large set of data and we need to find the fuzzy event matrices $\sigma_k, k = 1, 2, ..., L$. To align with the training of a typical neural network, we use the following network to approximate the cost function for our model, which can be expressed as:
$$
\begin{aligned}
Cost(\bar{q}, \bar{S})
    &= \frac{1}{2}||\bar{q_L}-\bar{q}||^2\\
    &= \frac{1}{2}\sigma^N_{n=1}(S^L_n - \bar{S_n})^2
\end{aligned}
$$
This series of operations, again, can be shown as the neural-network-resembled training architecture shown in figure 4.

With this series of equations ready, the rest of the work lie in determining the derivatives of the cost function for training the supervised learning algorithm using gradient descent, which is given by:
$$
\begin{aligned}
\bar{a}_{i,j}^k = -\gamma \frac{dCost}{da^k_{i,j}}
\end{aligned}
$$
Similar to most traditional supervised learning algorithms, $\gamma$ is a parameter while training.

For the output layer $L$, analogously, it can be determined as:
$$
\begin{aligned}
\frac{dCost}{da^L_{i,j}}
    &=\frac{dCost}{dS^L_j}\frac{dS^L_j}{da^L_{i,j}}\\
    &= (S^L_j-\bar{S}_j)\frac{dS^L_j}{da^L_{i,j}}\\
    &= (S^L_j-\bar{S}_j)\frac{d}{da^L_{i,j}}\frac{e^{\delta S^{L-1}_ia^L_{i,j}}}{\Sigma^N_{m=1}e^{\delta S^{L-1}_ma^L_{m,j}}}\\
\end{aligned}
$$
As a result of this calculation, we can determine that the error signal is:
$$\emptyset^L_j = S^L_j - \bar{S}_j$$.

Therefore, 
$$
\frac{dCost}{da^L_{i,j}}=\emptyset^L_jS^{L-1} \frac{e^{\delta S^{L-1}_ia^L_{i,j}}}{\Sigma^N_{m=1}e^{\delta S^{L-1}_ma^n_{m,j}}}
$$

Together, for layer, $n=1, 2, 3, ..., L-1$, we can similarly define the gradient descent, training step as
$$
\begin{aligned}
\frac{dCost}{da^k_{i,j}}
    &= \frac{dCost}{dS^k_j}\frac{dS^k_j}{da^k_{i,j}}\\
    &= (S^n_j-\bar{S}_j)\frac{d}{da^n_{i,j}}\frac{e^{\delta S^{n-1}_ia^n_{i,j}}}{\Sigma^N_{m=1}e^{\delta S^{n-1}_ma^n_{m,j}}}\\
\end{aligned}
$$

and error function as:
$$\emptyset^n_j = \frac{dCost}{dS^k_j}$$

Therefore,
$$
\frac{dCost}{da^n_{i,j}}=\emptyset^n_jS^{n-1} \frac{e^{\delta S^{n-1}_ia^n_{i,j}}}{\Sigma^N_{m=1}e^{\delta S^{n-1}_ma^n_{m,j}}}
$$

For the training purpose, we can recursively define the error signal, $\emptyset^k_j$ in the following form:
$$
\begin{aligned}
\emptyset^n_j 
    &= \frac{dCost}{dS^n_j}\\
    &= \Sigma^N_{k=1}\frac{dCost}{dS^{n+1}_k}\frac{dS^{n+1}_n}{dS^n_j}\\
    &= \Sigma^N_{k=1}\emptyset^{n+1}_k\frac{dS^{n+1}_k}{dS^n_j}\\
    &= \Sigma^N_{k=1}\emptyset^{n+1}_k a^{n+1}_{j,k}\frac{e^{\delta S^{n}_ja^{n-1}_{j,n}}}{\Sigma^N_{m=1}e^{\delta S^{n}_ma^{n+1}_{mk}}}\\
\end{aligned}
$$

With the adaptation algorithm above, the fuzzy states are calculated by $$S^n_j\approx \frac{1}{\delta}ln(e^{\delta S^{n-1}_ma^{k}_{m,j}}), n = 1, 2, ..., L$$.
And we can also calculate the back-propagation using the following:
$$
\begin{aligned}
\emptyset^L_j &= S^L_j - \bar{S}_j\\
\emptyset^n_j &= \Sigma^N_{k=1}\emptyset^{n+1}_k a^{n+1}_{j,k}\frac{e^{\delta S^{n}_ja^{n-1}_{j,n}}}{\Sigma^N_{m=1}e^{\delta S^{n}_ma^{n+1}_{mk}}}
\end{aligned}
$$
And gradient descent step is given by:
$$\bar{a}^k_{i,j} = -\gamma\emptyset^n_jS^{n-1}_i = \Sigma^N_{k=1}\emptyset^{n+1}_k a^{n+1}_{j,k}\frac{e^{\delta S^{n}_ja^{n-1}_{j,n}}}{\Sigma^N_{m=1}e^{\delta S^{n}_ma^{n+1}_{mk}}}$$

\subsubsection{Rolling-Window Regression:}
When training, rolling window regression is a simple yet power algorithm \citep{petropoulos2017using}. To make use of it in our case, we simply use the past 10 days (including today) to predict the very next day's stock price. This is a useful method, because of the Martingale theory that past prices are not as useful as the present price, see figure 5 below for how it works.
\begin{center}
    \includegraphics[width=0.5
    \textwidth]{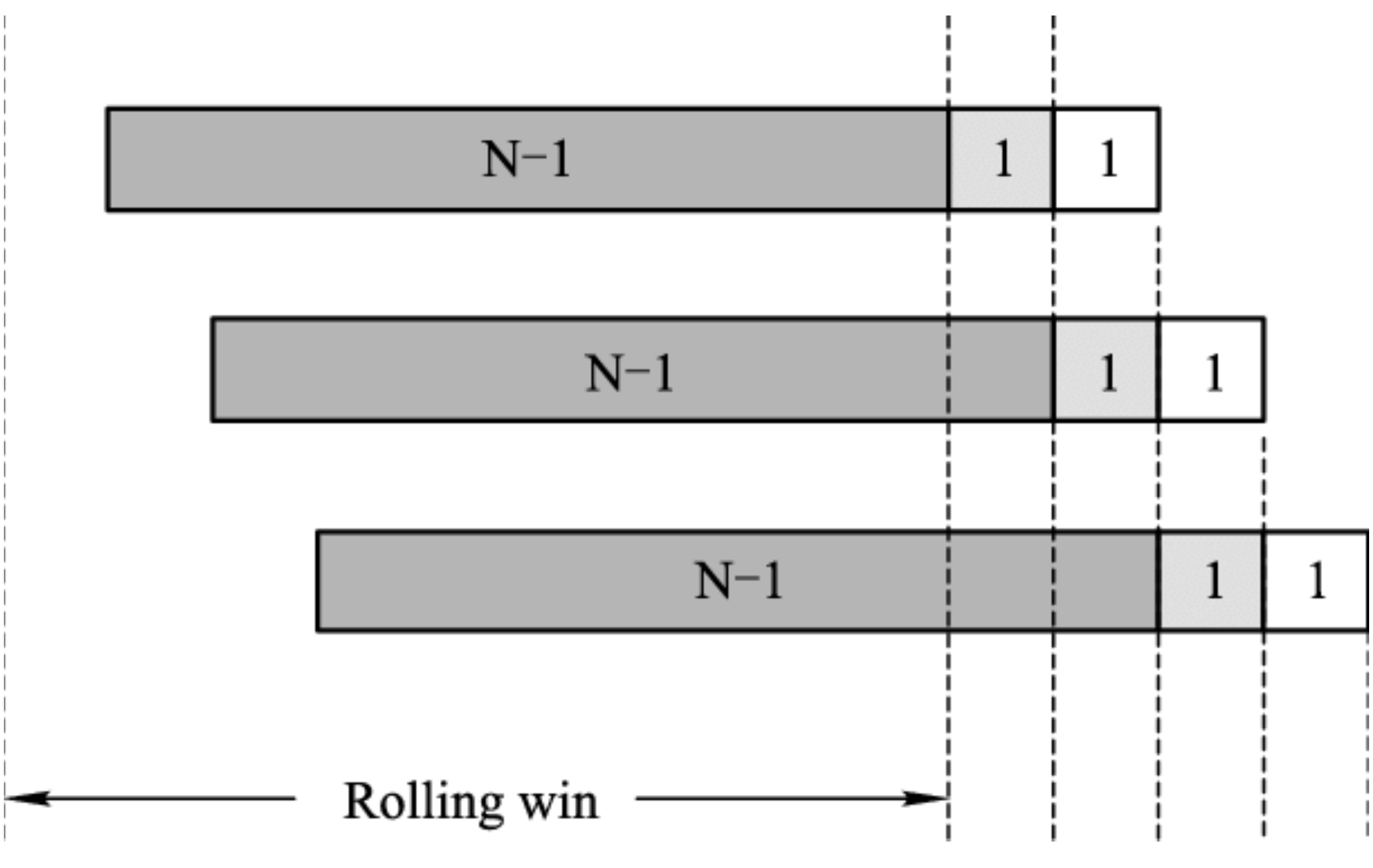}\\
    Figure 7: Rolling Window Regression
\end{center}
In this case, we want to give the most heavy weight to today's price as the basis for training, as diminishing weight for the days to the past from today. This training weight becomes zero once a time is before ten days before the presence. This is to say, any stock price from before 10 days ago are not used for the training of our model. This is also made possible by the real-time update system we have adopted. See section 2.2 regarding how we are combining the use of this algorithm with LSTM neural networks.
\section{RESULT}
The raw material industry appears to be one of the promising industries that we can apply our models to make predictions pretty well. This is because of the strong correlation between certain events, like weather, prices of raw materials, transportation and so on and the evaluation of raw material suppliers. So any event that changes any of the above parameters can cause quite significant impacts on the stock market of energy-related companies.

In the list of the 88 equities we looked into, some of the companies are such raw material suppliers. We will make an example out of Entegris, "ENGT."

For this example, if we simply apply a LSTM neural network to train for the stock price on the next time point based on the present value with the information adjusted by past stock price stretching out to a couple of days before the present, then the result will start accurate but grow more skewed to inaccurate predictions because of what the stock price of a specific time point is predicted based on. See figure 6 below and get a sense of how this would happen.

\begin{center}
    \includegraphics[width=0.6
    \textwidth]{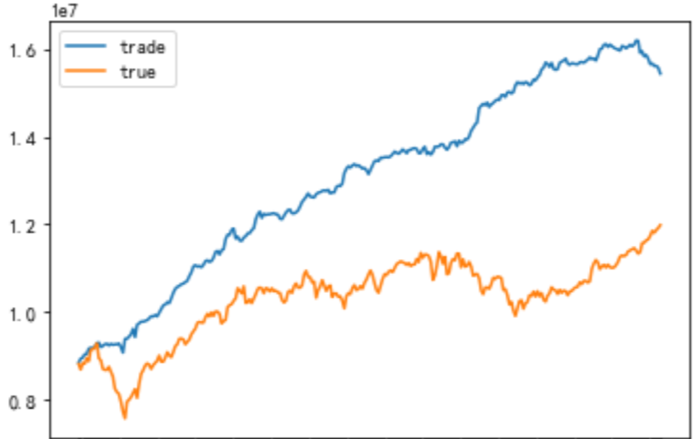}\\
    Figure 8: Traditional LSTM Network Result
\end{center}

Instead, another way of doing this, which the presented model does in this research, is instead to regenerate the impact of a certain event. Using these remodeled influences of certain macroeconomic events and adding it back on top of the traditional predictions, we can achieve a marginally better prediction result even in a more extended period of time.

We show our test of EB-dynaRA model on one of the raw material suppliers: Entegris, "ENGT." As previously mentioned, we firstly preprocess the graph so that we are only training on the time-series curve. We embed the new event-based supervised learning algorithm as the generator network to represent macroeconomic factors' influences on the stock. Then, once our GANs model is able to generate new stock trends that are real enough to fake the discriminator network, it means the new trends resemble the influence of the actual events well. Then, we add these adjustors to the traditional predictions based on LSTM Network. The result of a more extended period of time is as below:

\begin{center}
    \includegraphics[width=0.75
    \textwidth]{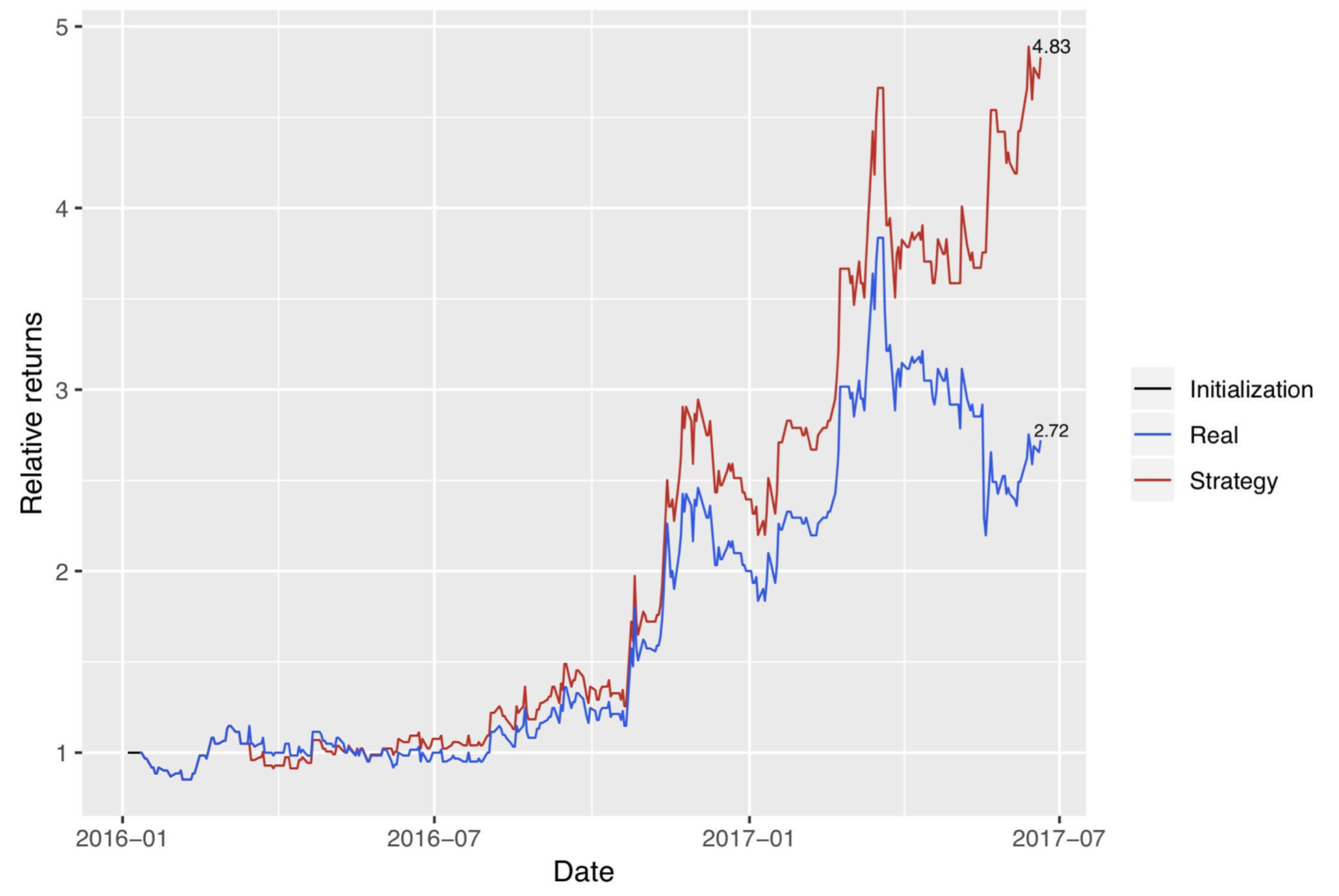}\\
    Figure 9: Adjusted LSTM Network Result
\end{center}

As shown, the prediction result on an even longer period of time shown in the Strategy curve is marginally closer to the Real curve and also mimic the pattern of the stock trend better at each turn, because of the incorporation of events as training parameters.
\section{DISCUSSION}
Event-based supervise learning algorithm is widely applicable, because training based on each real event is much representative of the real learning process and make how machines learn to do certain tasks much more transparent. Our application of it to predicting the changes in stock market is just one of many ways the new supervised learning algorithm based on fuzzy discrete event system could be used.

In addition, our treatment of the Brownian movement in the stock market also poses some possibilities of how to deal with other industries that involve this type of random movement. Brownian movement is know for its complexities of the influences of a mix of different forces. In our case, the noise of the stock market is formed by the complicated, interplexed impacts of a mix of the market, macroeconomic events. To formalize these events and quantify them into the form of vectors helps to organize the influence of these market forces. This method of regenerating a mix of forces into one adjuster that resemble the effects of a large number of forces/events that are hard to specify facilitate the process.

With this model of building adjusters having been proposed, it is also important to realize the practicality and importance of putting together a database/library of the effects of different events. For example, if there is an outbreak of virus in a certain size of area, impacting a certain number of people, how would this effect the market? Once we could formalize this process of determining the event matrix for this specific events, more often than not, these matrices are directly applicable to other time/circumstances similar events happen. With a database built, the entire process of making the adjuster lie in putting together the influence of different events that happen on a given time and use them to produce an adjuster. This adjuster is capable of making the traditional predictions purely based on a statistical approach to find fit more attached to the true world. This process could be considered similar to how \textit{a posteriori} would impact the judgement of \textit{a priori}.
\section{CONCLUSION}
This novel event-based supervised learning is potentially game-changing. Its application on predicting the changes in the stock market based on real-world macroeconomic events shows it significant potential as a way to adjust for innate statistical predictions that purely make predictions based on trends itself. In this research, we focus on an event-based system's potential on the quantitative finance industry. We first provided an overview of different ways of making stock market predictions, traditional or the more recent quantitative computational approaches. Though all of them have achieved some of the most cutting-edge results, some assumptions about the market and predicting the market are made so that many of the methods are restricted toward directly looking at the pattern of the time-series graph rather than incorporating specific events into the prediction. In this research, we used our novel event-based training model along with some other traditional models, including LSTM and GANs, to produce some of the state-of-the-art results in the industry. However, it is still important that more event matrices are tested and built in the future and form a database for the convenience of applying them with the event-based model proposed in this research. We believe that the idea of incorporating events as a part of the training can potentially push the tides of many different industries that involve pattern recognition as a part of their strategies.
\section{ACKNOWLEDGEMENT}
We would like to thank Dr. Feng Lin, an IEEE Fellow at Wayne State University, for his consultation on this project. He helped explain and develop some key concepts of the supervised learning algorithm based on Fuzzy Discrete Event System (see section 3.3) in this research.

\bibliographystyle{plainnat}  

\bibliography{ebdynara}

\end{document}